# Impact of Brain Anisotropy on Transcranial Temporal Interference Stimulation: Numerical Analysis Toward Reliable Montage Optimization

Kanata Yatsuda[1], Masaki Fukunaga[2,3,4], Wenwei Yu[1,5], Jose Gomez-Tames[1,5]*

[1]Department of Medical Engineering, Graduate School of Engineering, Chiba University, Chiba 263-8522, Japan

[2]Section of Brain Function Information, National Institute for Physiological Sciences, Okazaki 444-8585, Japan

[3]Physiological Science Program, Graduate Institute for Advanced Studies, SOKENDAI, Hayama 240-0193, Japan

[4]Core for Spin Life Sciences, Okazaki Collaborative Platform, National Institutes of Natural Sciences, Okazaki 444-8585, Japan

[5]Center for Frontier Medical Engineering, Chiba University, Chiba 263-8522, Japan

* Corresponding author:
Jose Gomez-Tames
jgomez@chiba-u.jp

## Abstract

**Background & Aim:** Transcranial temporal interference stimulation (TIS) is a novel transcranial electrical stimulation modality that enables focused targeting of deep brain structures. When targeting deep regions, current pathways traverse the highly anisotropic white matter, making anisotropy a potentially critical factor. This study aimed to clarify how anisotropy influences interferential currents in deep brain regions and to assess its impact on TIS montage optimization for the first time.

**Methods:** Anatomical head conductor models with anisotropic and isotropic conductivities were compared to evaluate the role of anisotropy in the intracranial interferential currents distributions and montage optimization. For the anisotropic conductivity, conductivity tensors were derived from diffusion-weighted imaging data for gray matter, white matter, and deep brain structures. Montage optimization was conducted based on Pareto front optimization.

**Results:** In literature-reported TIS montages, anisotropic conductivity significantly altered the interferential electric field intensity, with differences of up to 18% in the white matter, whereas discrepancies in deep brain structures remained below 12%. For TIS-optimized montages, these variations across conductivity models yielded only 50% montage disagreement. However, when constraining differences in focality and target field strength to within 10%, the agreement improved to nearly 90%.

**Conclusions:** Incorporating anisotropic conductivities is important to determine individualized focality and target-field characteristics. Moreover, anisotropy can substantially affect the selection of the optimal montage, but under practical focality and target-field tolerances, montage choices largely converge.

**Significance:** For practical montage optimization, isotropic conductivity often suffices within typical tolerance bounds, enabling planning when diffusion data are unavailable. Anisotropy remains essential for mechanistic interpretation, orientation-dependent hypotheses, and patient-specific predictions where white-matter effects are critical.

**Introduction**

Transcranial electrical stimulation (tES) is a non-invasive technique that delivers a weak electric current through scalp electrodes [1], [2]. The resulting electric field modulates cortical regions involved in higher cognitive functions and motor control [3]. Currents can penetrate beyond the cortex and modulate subcortical structures, reaching up to 70% of cortical intensities with conventional bipolar montages [4]. Consequently, the use of tES for deep modulation has been of interest, but it is associated with low focality and a risk of unintended effects. One approach is multi-electrode tES, which can improve focality while minimizing cortical intensities [5]. However, increasing montage complexity limits patient burden and clinical scalability, motivating the development of alternatives that shift effective modulation to depth with fewer electrodes.

Transcranial temporal interference stimulation (TIS) is a tES modality proposed to target deep brain structures while limiting superficial co-stimulation [6]. TIS delivers two slightly different kilohertz currents (e.g., 2000 Hz and 2010 Hz) through two scalp electrode pairs [6] (Figure 1). Their superposition yields an amplitude-modulated signal whose envelope oscillates at the beat frequency (e.g., 10 Hz). The envelope's modulation depth (envelope peak-to-peak amplitude) is the putative driver of neural modulation [6], [7]. By optimizing the TIS montage, high modulation depth can be directed to a deep target while remaining low elsewhere, where predominantly high-frequency components persist [6], [8].

TIS has demonstrated deep, spatially focused stimulation in humans [9], [10], [11], yet other studies report null or small effects at depth and unintended off-target stimulation [12], [13]. These inconsistencies likely reflect the difficulty of montage selection for steering interferential current, considering individual variations and complex brain anatomy [14], [15]. Current montage selection relies on computational electric-field modeling with subject-specific head models that incorporate realistic anatomy and tissue conductivity [14], [16], [17], [18]. Most tES simulations assume isotropic conductivity [19], as various studies find only modest differences in the estimated cortical fields relative to anisotropic conductivity [20] [21], [22]. However, when targeting deep brain structures, current pathways pass through highly anisotropic white matter, making anisotropy non-negligible. To address this, some studies incorporate diffusion-weighted imaging (DWI)-derived anisotropic conductivity in TIS [13], [23], [24]. However, there is currently no consensus regarding the adoption of DTI-based conductivity in computational modeling, particularly for interferential stimulation. In addition, DWI adds computational cost and requires additional scanning time. Therefore, it is essential to determine how anisotropy shapes interferential currents at depth and how it influences TIS montage optimization.

The aim of this study is to investigate the effect of anisotropy on TIS montage optimization for deep brain targets and to provide a rationale for whether its inclusion in computational simulation is necessary. First, we examine how the electric field differs between isotropic and anisotropic models. Second, we evaluate how these differences influence the selection and performance of the optimal montage, weighing the costs and benefits of adopting anisotropic modelling as standard practice.

**Model and Methods**

Both the experimental and computational components of this study were reviewed and approved by the Ethics Committees of Chiba University (approval code: R6-33) and National Institute of Natural Sciences (approval code: EC01-098), and were conducted in accordance with the principles outlined in the Declaration of Helsinki.

**Head and Electrode Models**

Individual head models were generated for fifteen young subjects (12 males, 3 females; 21.9 ± 1.6 years) from MRI-T1 image (TR = 2400 ms, TE = 2.24 ms, TI = 1060 ms, flip angle = 8°, FOV = 240 × 256 mm², matrix = 300 × 320, voxel size = 0.8 × 0.8 × 0.8 mm³), MRI-T2 (TR = 3200 ms, TE = 560 ms, flip angle = 120°, FOV = 240 × 256 mm², matrix = 302p × 320, voxel size = 0.8 × 0.8 × 0.8 mm³), and diffusion-weighted MR images (TR = 4000 ms, TE = 90 ms, GRAPPA factor = 2, Multi-band acceleration factor = 3, FOV = 200 × 200 mm², matrix = 118p × 118, voxel size = 2.0 × 2.0 × 2.0 mm³). Segmentation was conducted using SIMNIBS/charm [25]. These models comprised nine tissues: scalp, compact bone, spongy bone, blood, muscle, eyeballs, cerebrospinal fluid (CSF), gray matter, and white matter. Deep brain structures were segmented using FreeSurfer/recon-all (ver. 7.4.1) and incorporated into the head model. A total of eight structures were included: the thalamus, putamen, caudate, pallidus, hippocampus, amygdala, accumbens, and ventral DC, as shown in Figure 2. Electrodes were placed according to the 10-10 system and consisted of a 4-mm-thick rubber sheet (conductivity of 1.0 S/m) with a diameter of 1.5 cm [26], [27].

**Isotropic and Anisotropic Conductivity**

Two conductivity models were compared in TIS: an isotropic scalar model and a volume-normalized anisotropic model.

In the isotropic model, each tissue was assigned uniform scalar conductivities $\sigma_{iso}$ (S/m) representing the conventional simplification commonly adopted in tES simulations: scalp (0.456), compact bone (0.008), spongy bone (0.025), blood (0.6), muscle (0.16), eyeballs (0.5), CSF (1.654), gray matter (0.275), and white matter (0.126). [28], [29], [30]. The isotropic conductivity of the deep brain structures was assumed to be the same as that of gray matter [4].

In the anisotropic model [22], conductivity tensors were derived from DWI data for gray matter, white matter, and deep brain structures via the SimNIBS/dwi2cond [31], [32]. Briefly, the diffusion tensor at each voxel was decomposed into three eigenvalues, representing diffusivity magnitudes, and corresponding eigenvectors, representing diffusion orientations. To ensure fair comparisons across voxels, we applied volume normalization to the anisotropic eigenvalues. Direct use of diffusion eigenvalues would result in voxel-wise variability in mean conductivity. Volume normalization removed these differences by equalizing the average conductivity across voxels, allowing anisotropy to be evaluated independently of absolute conductivity values. The normalized eigenvalues and their eigenvectors were then recombined to reconstruct anisotropic conductivity tensors aligned with subject-specific white matter fiber orientations and scaled consistently with isotropic conductivity references.

**Electric Field Model and Lead Field**

Electric fields were computed using the Laplace equation via a finite element method solver (SimNIBS 4.0 [33], [34]) on subject-specific head models. TIS was modeled by the superposition of two fields ($\vec{E}_1$ and $\vec{E}_2$) that correspond to each of the two electrode pairs. The resulting maximum envelope's modulation depth was obtained using the formulation proposed by Grossman et al. [6]:

$$|\vec{E}_{AM}(\vec{r})| = \begin{cases} 2|\vec{E}_2(\vec{r})| & if\ |\vec{E}_2(\vec{r})| < |\vec{E}_1(\vec{r})|cos\alpha \\ 2\left|\vec{E}_2(\vec{r}) \times \left(\vec{E}_1(\vec{r}) - \vec{E}_2(\vec{r})\right)/|\vec{E}_1(\vec{r}) - \vec{E}_2(\vec{r})|\right| & otherwise \end{cases} \quad (1)$$

The $\vec{E}_1(\vec{r})$ and $\vec{E}_2(\vec{r})$ correspond to the electric fields at position $\vec{r}$. If $|\vec{E_1}| > |\vec{E_2}|$ and the angle $\alpha$ between $|\vec{E_1}|$ and $|\vec{E_2}|$ is smaller than 90°, the maximal modulation depth is

obtained. If $\alpha$ is more than 90°, the sign must be reversed. Simulations were performed for both isotropic and volume-normalized anisotropic conductivity models to assess differences in field distribution. To eliminate potential computational artifacts, the 99.9th percentile values of $\vec{E}_1$ and $\vec{E}_2$ in each brain tissue were excluded from the analysis [35].

To efficiently evaluate a large number of electrode montages, we implemented a lead-field matrix [36]. One electrode was designated as the fixed return, and a current of 1 mA was sequentially applied to each of the remaining $N-1$ electrodes. The resulting fields were computed and stored. Any TIS montage, defined by two electrode pairs within the $N$ set, was then obtained as a weighted linear combination of these fields, with weights corresponding to the applied currents. This reduced computation time from about two minutes per FEM simulation to under 0.1 seconds. The full matrix for $N=72$ electrode sites can be generated in approximately one hour, making this method highly suitable for large-scale optimization.

**Reference TIS Montages**

To compare the TIS electric field distributions between isotropic and anisotropic conductivities, we used TI montages previously reported in the literature for various deep brain targets. Montage 1 (AFz–P7, AF4-P9, [14]) was optimized for the hippocampus by maximizing the median electric field strength within the target while minimizing the volume of other brain tissues exposed to fields exceeding 0.25 V/m. Montage 2 (P8-TP7, Fp2-FT7, [9]) was validated for the left hippocampus, with simulations showing that the median envelope modulation amplitude in the hippocampus was 30–60% greater than in the overlying cortex. Montage 3 (F7-PO7, F8-PO8, [18]) targeted the thalamus using a bilaterally symmetrical configuration designed to maximize the electric field in the thalamus while minimizing the electric fields in surrounding tissues, based on visual inspection. Montage 4 (F3-F4, TP7-TP8, [10]) targeted the left putamen using a multi-objective optimization that maximizes fields in target, a focality ratio to ensure selectivity, and an activation ratio to ensure adequate target coverage. Montage 5 (F6-F7, PO10-F5, [37]) was optimized to the left Putamen by maximizing the average electric field in the target and the ratio of the average electric field in the target region relative to gray matter.

**Optimization of TIS Montages**

The optimal montage was determined for each deep brain target in each individual under both isotropic and anisotropic conductivity conditions. Quasi-optimization was based on the trade-off between the average electric field in the target region ($EF_{ROI}$) and the focality ratio, defined as the average electric field in the target region relative to that in the gray matter ($ROI_{Focality} = EF_{ROI}/EF_{GM}$) [37]. As shown in Figure 3A, this trade-off forms a Pareto front, and the montage with the shortest distance to the ideal point was selected as optimal. The search space consisted of 200,000 randomly generated TIS montages, drawn from 72 candidate electrode locations based on the International 10–10 system, excluding the positions over the ears.

**Evaluation Metrics**

For the montages reported in the literature, we quantified the discrepancy between electric field distributions obtained with anisotropic and isotropic conductivities ($EF_{ani}$ and $EF_{iso}$) using two complementary metrics: relative $L^2$-norm error and the pointwise relative error.

The relative $L^2$-norm error quantifies the global root-mean-square deviation of $EF_{ani}$ and $EF_{iso}$, normalized by the overall magnitude of $EF_{ani}$. This metric penalizes large local deviations. For electric fields, which often exhibit spatially concentrated high-intensity regions, the relative $L^2$-norm emphasizes errors in these dominant regions.

$$Rel\ L^2 = \frac{\sqrt{\sum_{i=1}^{N}(EF_{ani_i} - EF_{iso_i})^2}}{\sqrt{\sum_{i=1}^{N}(EF_{ani_i})^2}} \quad (2)$$

We also computed the pointwise relative errors at each spatial location, including their median and maximum values. 99.9th percentile of $e_i^{rel}$ in each brain tissue were removed to minimize potential metric artifacts. The median reflects the typical local deviation, and the maximum identifies the worst-case error. This metric is particularly relevant for electric fields where both high- and low-intensity regions are of interest.

$$e_i^{rel} = \frac{|EF_{ani_i} - EF_{iso_i}|}{|EF_{ani_i}|} \quad (3)$$

For montages optimized under anisotropic or isotropic conditions, discrepancies were evaluated by comparing their performance on the $EF_{ROI}$ vs. $ROI_{Focality}$ graph. A tolerance region, centered on the anisotropic-based optimal montage (Figure 3B), was defined to assess similarity. When isotropic-based montages produced results within the tolerance region, the two montages were considered comparable. The size of the tolerance circular region varied from 0% to 30% in increments of 0.01%. This analysis emphasized performance outcomes rather than electrode position difference, since distinct electrode configurations can yield similar performance [37].

**Results**

**Electric Field Variation between Anisotropic and Isotropic Conductivities**

Figure 4 shows the cross-section of group-level fields for anisotropic and isotropic conductivities, using reported TIS montages from the literature across all subjects. The differences are particularly pronounced in the white matter, and this tendency was consistently observed across all individuals, in particular for Montages 2, 4, and 5. Gray matter exhibits minimal global differences in the electric field across all montages. For deep brain tissues, some differences were observed, particularly the superior bilateral portions that are in contact with white matter.

We quantified the differences in intracranial electric fields due to anisotropic conductivities in Figure 5. In the case of the gray matter, the relative $L^2$-norm error reaches up to 9%, and pointwise relative errors are up to 5% and 53% for median and maximum errors, respectively, among the montages. In the case of the white matter, the relative $L^2$-norm error reaches up to 18%, and pointwise relative errors are up to 9% and 49% for median and maximum errors, respectively, among the montages. Across the deep targets, the relative $L^2$-norm error reaches up to 17%, and pointwise relative errors are up to 11% and 66% for median and maximum errors, respectively, among the montages. The putamen (Montages 4 and 5) presents the higher differences (relative $L^2$-norm: 13%, maximum error: 48%). It is followed by the thalamus (Montage 3; $L^2$-norm: 8%, maximum error: 24%) and hippocampus (Montages 1 and 2; relative $L^2$-norm: 8%, maximum error: 33%). Inter-subject variability was generally low. For all metrics except maximum error, the standard deviation was below 1.6 ± 0.8 %, indicating low variability across subjects. Even for the maximum error, the variability remained below 4.4 ± 2.0 %.

**Impact of Anisotropic Conductivity on Optimal Configurations**

Figure 6A shows the difference in performance between montages optimized under isotropic and anisotropic conditions. The putamen exhibited the largest increase under anisotropic conductivity, with both metrics rising by approximately 10%. In contrast, the hippocampus showed a decrease in both metrics under anisotropic conditions. When comparing the left and right hemispheres of the same target, the differences were up to 5% with for the amygdala in terms of $EF_{ROI}$ and $ROI_{Focality}$, respectively.

Figure 6B shows the match rate between isotropic-based and anisotropic-based montages' performance at different tolerance regions (see Figure 3B). The match rate was obtained among 255 combinations (15 participants × 17 targets). The results demonstrate a gradual increase in the match rate, with values of 50.1 ± 4.7% at 0% tolerance ratio (identical performance), 87.8 ± 1.8% at 10%, 97.7 ± 1.0% at 20%, and 99.8 ± 0.4% at 30%. The standard deviations were consistently low (<5%), indicating high stability across repetitions.

Figure 6C shows the match rate per target. At 0% tolerance ratio, the match rate varied widely across targets and did not reach 60% on average. At a 10% tolerance ratio, the match rate exceeded 80% for all targets. When the tolerance increased to 20%, the match rate exceeded 96% for all targets except the left Thalamus, which reached 92%.

**Discussion**

The distribution of electric fields can be strongly influenced by the anisotropic nature of the white matter [38], [39]. However, its impact has not been quantified in TIS studies, particularly when targeting deep brain structures, such as TIS, [14], [15]. Moreover, the effect on estimating the montage is unknown.

*Impact on Electric Field Distribution*

We evaluated the impact on interferential electric field distributions when incorporating anisotropic conductivities using montages previously reported in the literature. In general, gray matter presented a $Rel\ L^2$ within 10%. The white matter presented larger differences up to 18% for $Rel\ L^2$. In general, deep brain tissues also had a $Rel\ L^2$ up to 17%.

Our results suggest that the electric field distribution in deep brain regions is mainly influenced by the anisotropy of surrounding white matter, such as the corpus callosum and internal capsule. These tracts have dominant fiber orientations: left–right in the corpus callosum and superior–inferior in the internal capsule [40]. When the electric field is aligned with these directions, anisotropy can strongly distort the field. In contrast, placing the field orthogonal to these fibers may reduce such effects. Notably, when targeting the left hippocampus and putamen, the degree of impact on these tissues varied by electrode configuration, with montages 2, 4, and 5 exerting the greatest influence.

Among the tested configurations, Montage 1 and 3 showed the smallest impact from anisotropic conductivity, which used a bilateral posterior-anterior alignment. In montage 3, it is possible to minimize left–right oriented interference fields, as such fields interact strongly with major fiber directions of the corpus callosum. However, a symmetric electrode configuration may reduce targeting accuracy, especially for unilateral deep targets. These findings highlight the importance of considering white matter anisotropy when designing stimulation strategies for deep-brain targets.

In the case of pair-wise error analysis, the median relative error exhibits values of 5–11% across all tissues, with higher values in white matter and some deep targets. The maximum relative error exceeded 29% in every tissue and reached 66%, indicating that anisotropy can induce large deviations at the voxel level, even when central tendencies are modest. Because the Pareto front was constructed from mean fields within each target tissue, these localized outliers have limited leverage on the optimization.

*Impact on Montage Selection*

Comparing isotropic and anisotropic field maps alone does not reveal the impact on montage selection. We therefore introduce a performance-based similarity metric with a predefined tolerance band; montages that achieve comparable target-field strength and focality are considered equivalent. This metric quantifies the extent to which anisotropy changes the functional effect of the chosen montage, rather than whether its scalp coordinates are identical.

With a 10% tolerance band, 90% of montages matched between isotropic and anisotropic models; with zero tolerance (identical montage), the match rate was 50%. At a 20% tolerance, the montage agreement reached 96% across all targets and subjects, except for the left thalamus. Although focality and field magnitude vary across deep targets, these differences have little influence on the optimal montage (Fig. 6C). Practically, this supports optimizing electrode placement with isotropic conductivities when detailed anisotropy is unavailable. However, when interpreting stimulation effects from the chosen montage, field orientation matters: when the induced field aligns superior–inferior, anisotropy exerts a smaller influence and isotropic and anisotropic predictions converge.

*Limitations and Future Work*

The current study has various limitations. First, modeling approaches typically assume subject-invariant tissue conductivities [42], [43]. Sensitivity analysis accounts for 20% of the variations in peak electric field values [44]. Second, the cohort comprised young adults; age-related cranial calcification can alter skull conductivity and primarily affects field magnitude, with smaller effects on spatial patterns [45][46]. Additionally, a DTI study comparing young and elderly populations has shown age-related decreases in FA. It increases in MD in white matter, but increases in both metrics in deep brain structures, such as the putamen and caudate [47]. These findings highlight the need for studies across a broader age range to gain a better understanding of age-related tissue changes and their impact on stimulation outcomes. Finally, this study investigated the modulation depth along the direction yielding maximum modulation. Future work should refine this by quantifying anisotropy–orientation coupling along relevant pathways [48][49] and apply tract-informed optimization to select montages that exploit these properties.

**Conclusion**

We investigated two questions: (i) how tissue anisotropy reshapes intracranial interferential electric fields (i.e., modulation magnitude and distribution), and (ii) how those anisotropy-driven changes alter the selection of the "optimal" electrode montage with clinical implications. Incorporating anisotropic conductivities changed field magnitudes by 18% in white matter and 6–17% across deep structures, depending on current orientation (montage). These shifts resulted in a 50% disagreement in the chosen montage using anisotropic and isotropic models. However, when requiring both focality and target-site field strength to remain within 10% of design goals, disagreement fell to 10% (90% agreement). Thus, anisotropy is relevant for determining the intracranial interferential currents modulation intensities, but under reasonable targeting tolerances, its impact on final montage choice is limited.

**Disclosure**

This study was supported by a JSPS Grant-in-Aid for Scientific Research, JSPS KAKENHI Grant- 25K15887, JSPS Program for Forming Japan's Peak Research Universities（J-PEAKS）Grant Number JPJS00420230002, and the Joint Research Program (25NIPS632) of the National Institute for Physiological Sciences, MEXT/CURE JPMXP1323015488 (Spin-L program No, spin25XN050).

**Conflict of Interest**

The authors declare that they have no conflicts of interest.

**Figures**

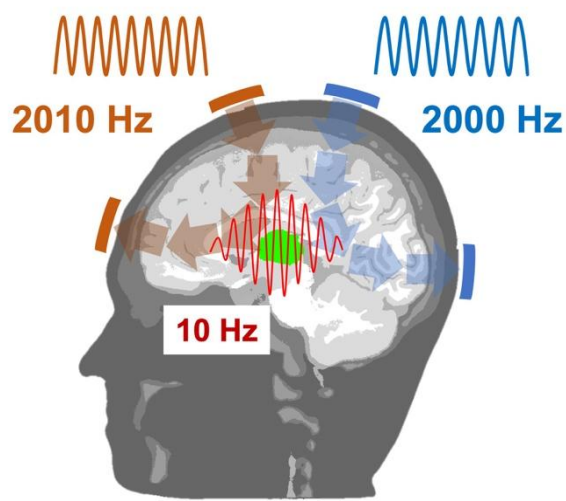

**Figure 1.** Transcranial temporal interference stimulation (TIS) injects two kilohertz currents of slightly different frequency via two pairs of electrodes that produces an amplitude-modulated field pattern $\vec{E}_{AM}$.

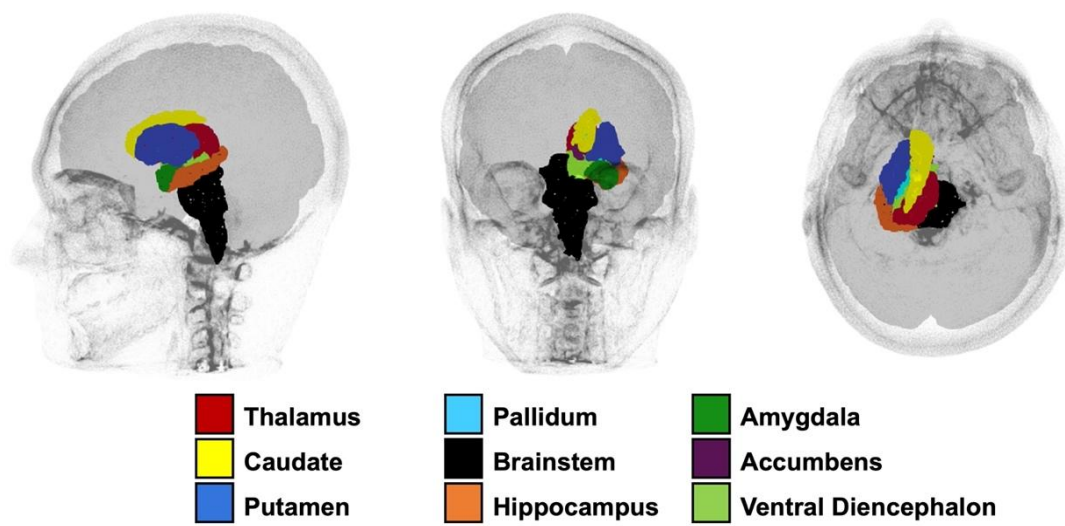

**Figure 2.** Deep brain targets.

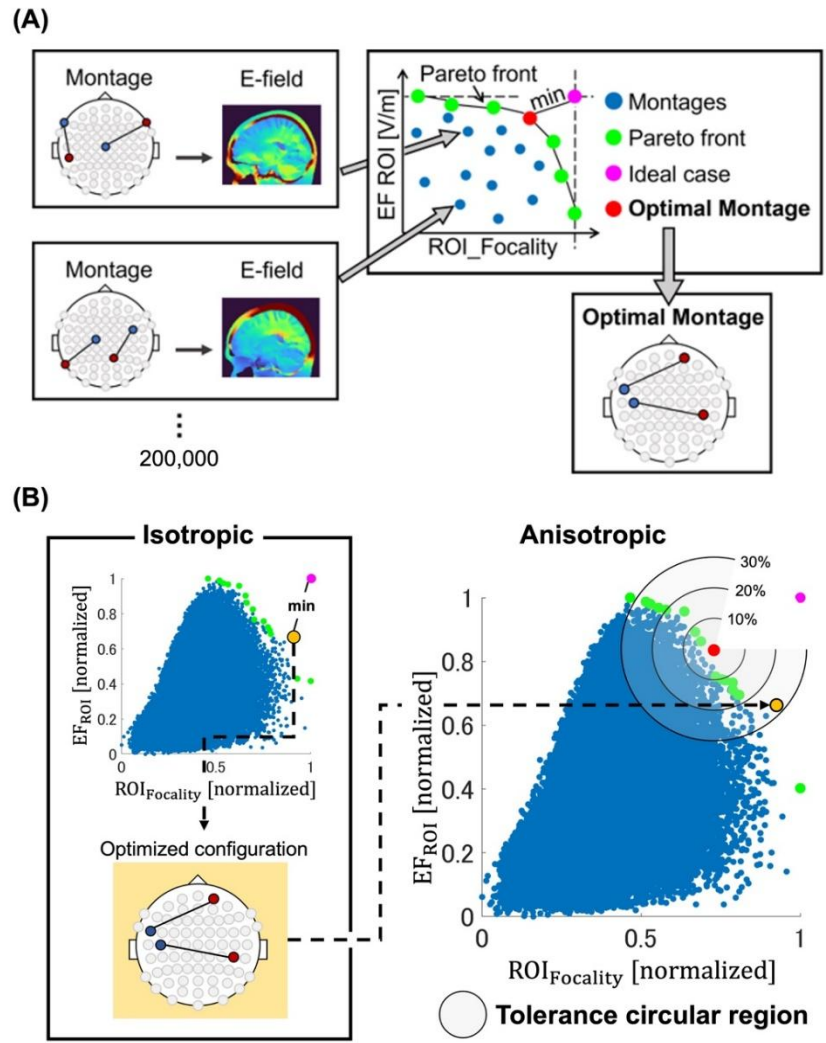

**Figure 3.** (A) Optimization method. TIS electrode montage is optimized by maximizing the trade-off between the average electric field in the target region ($EF_{ROI}$) and the ratio of the average electric field in the target region to that in the gray matter ($ROI_{Focality} = EF_{ROI}/EF_{GM}$). The montage with the minimum distance to ideal one was chosen to maximize $EF_{ROI}$ and minimize $ROI_{Focality}$ (i.e., minimizing unintended stimulation in the cortex). (B) Comparison of the performance of the optimal electrode configurations under isotropic and anisotropic conductivities. The circular region indicates a tolerance area centered on the optimal configuration derived from the Pareto front under anisotropic conductivity. If the performance of the optimal montage under isotropic conductivity is inside the tolerance area, it is considered that there is no difference in terms of montage optimization under both conductivity conditions. Tolerance areas are evaluated from 10% to 30%.

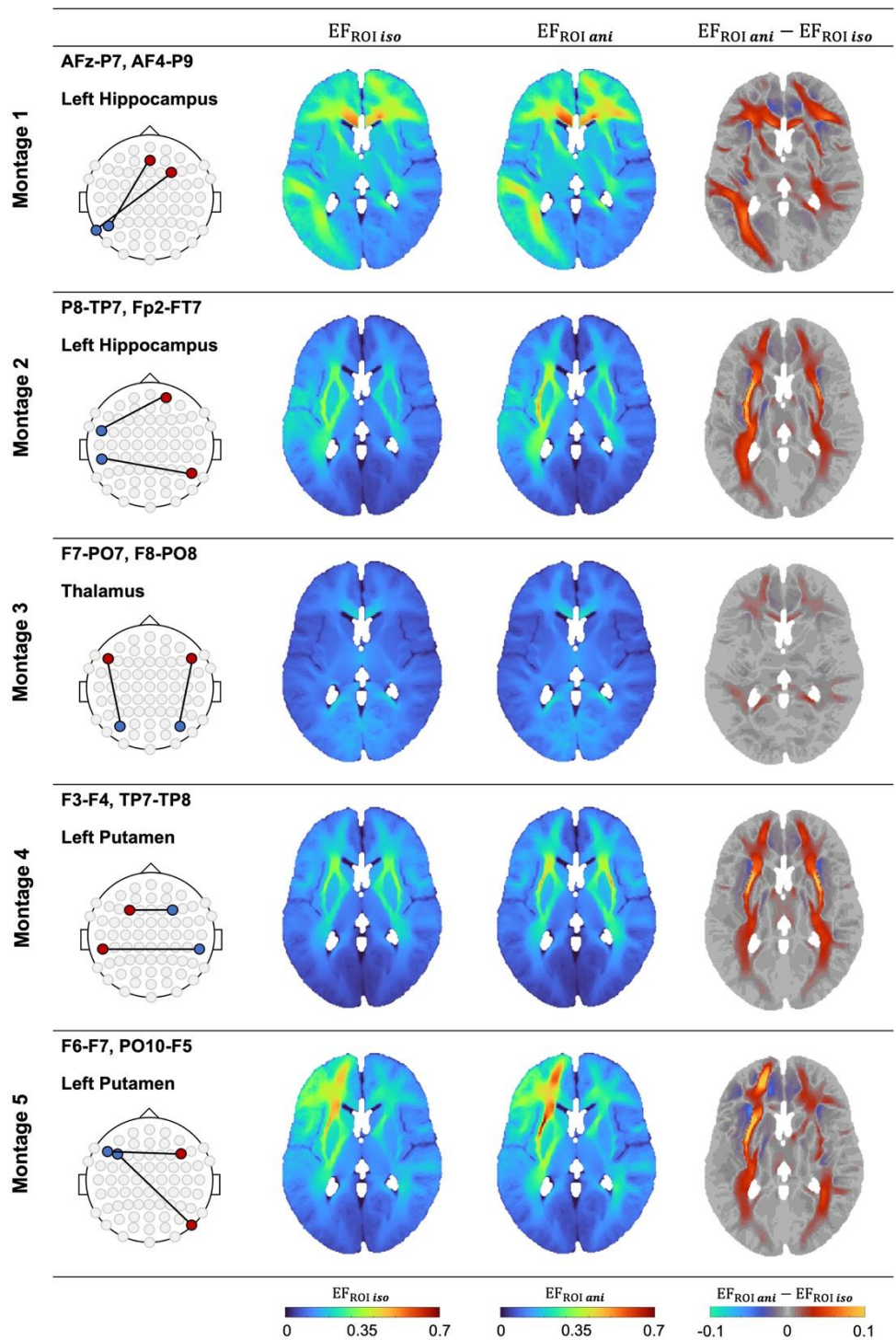

**Figure 4.** Group-level interferential electric fields shown in the MNI152 template using electrode configurations reported in previous studies for 15 subjects.

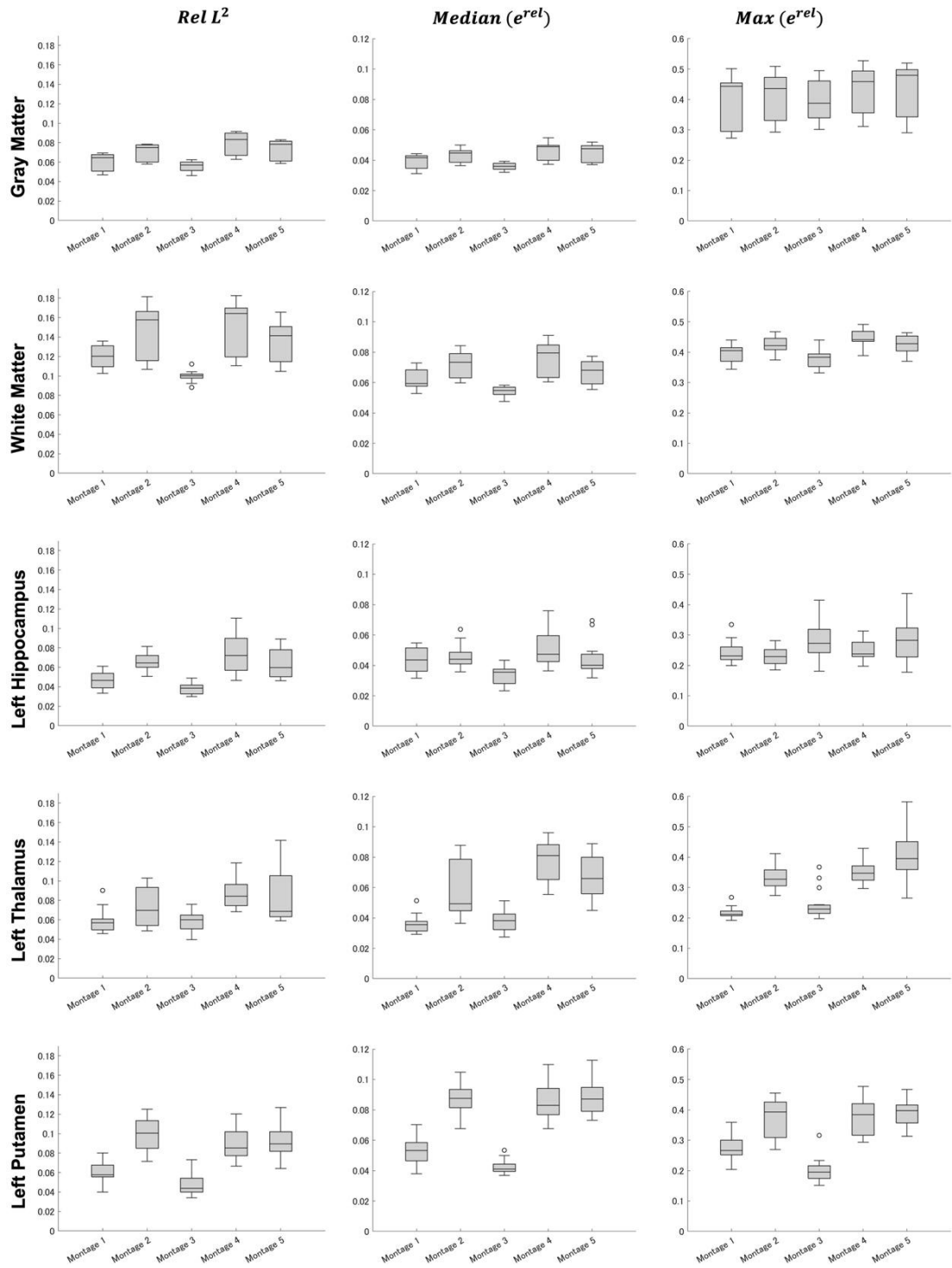

**Figure 5.** Error metrics represent the differences in electric fields between isotropic and anisotropic conductivity models in each deep brain structure ($n$ = 15).

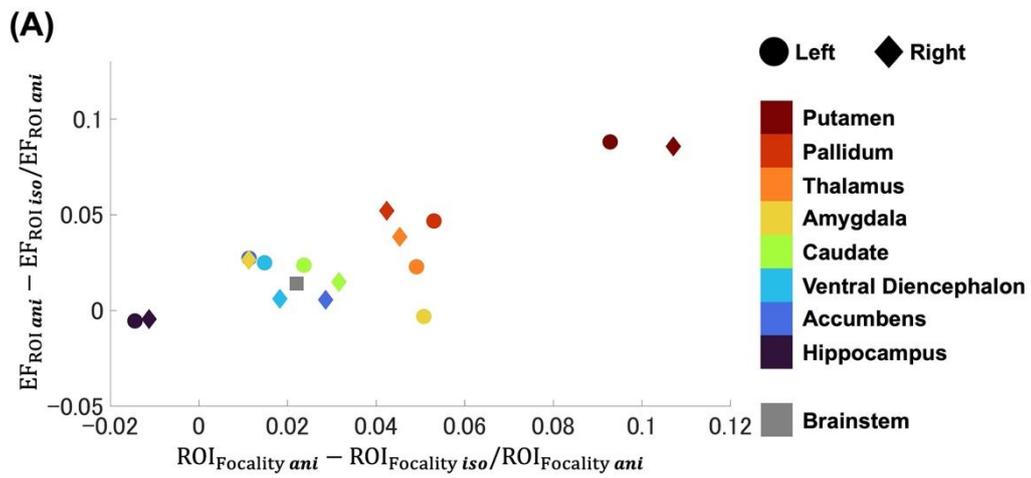

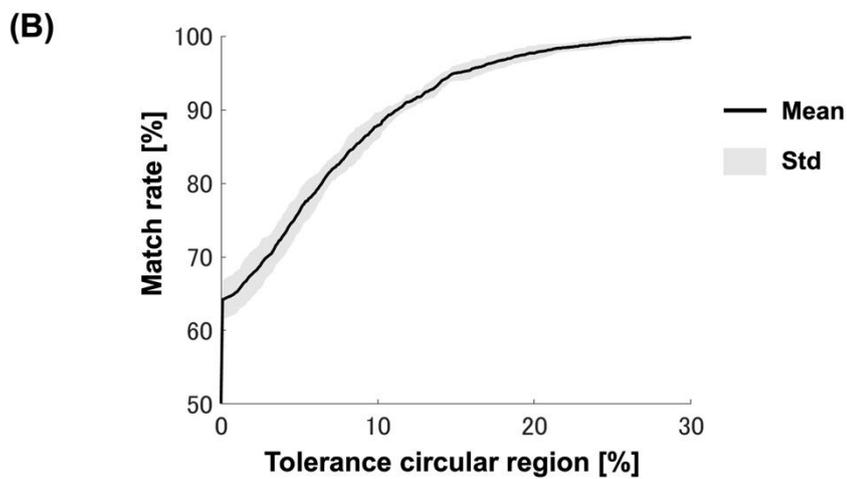

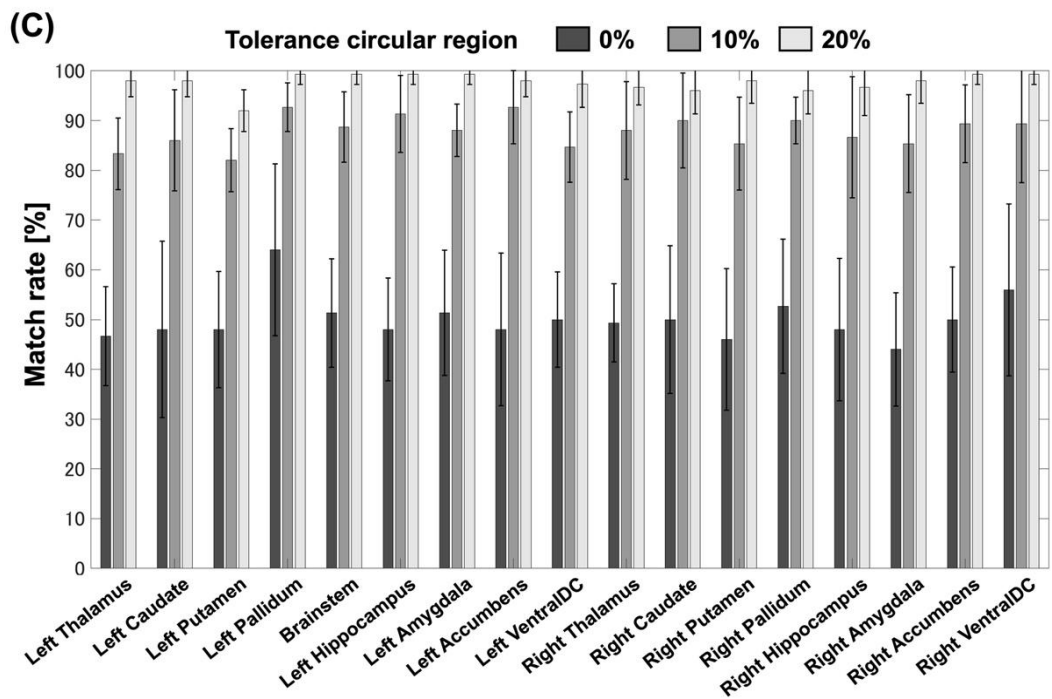

**Figure 6.** Comparison of optimal electrode configurations under isotropic and anisotropic conductivity. (A) Average differences in two performance metrics—target field strength and focality—between isotropic and anisotropic conductivity, calculated for each target. (B) A tolerance circular region was defined around the anisotropic-based optimal montage. If the corresponding isotropic-based optimized montage fell within this region, it was considered comparable. The radius was varied from 0% to 30% in 0.01% steps. The black line indicates the match rate calculated from 255 cases (17 targets × 15 head models), with mean and standard deviation obtained across 10 repetitions (search on 10 different random subspaces of 200,000 montages). (C) Match rate was calculated separately for each target across 15 head models. Tolerance circular regions were set at 0%, 10%, and 20%, and the corresponding match rates are shown for each target.